\documentclass{ws-acs}
\usepackage[latin9]{inputenc}
\usepackage{prettyref}
\usepackage{textcomp}
\usepackage{amstext}
\usepackage{graphicx}

\makeatletter

\newcommand{\lyxmathsym}[1]{\ifmmode\begingroup\def\b@ld{bold}
  \text{\ifx\math@version\b@ld\bfseries\fi#1}\endgroup\else#1\fi}

\newcommand{\lyxaddress}[1]{
\par {\raggedright #1
\vspace{1.4em}
\noindent\par}
}

\newcounter{myctr}

\usepackage{url}

\makeatother

\begin{document}
\makeatletter \global\long\def\@#1#2#3#4#5{[#1]}
 \makeatother

\markboth{Crampes, Plantié}{A Unified Community Detection, Visualization
and Analysis method}

\catchline{}{}{}{}{} 

\title{A Unified Community Detection, Visualization and Analysis method }

\author{Michel Crampes}

\author{Michel Plantié}

\maketitle

\lyxaddress{Ecole des Mines d'Ales, Parc Georges Besse, 30035 Nîmes Cedex }
\begin{abstract}
With the widespread of social networks on the Internet, community
detection in social graphs has recently become an important research
domain. Interest was initially limited to unipartite graph inputs
and partitioned community outputs. More recently bipartite graphs,
directed graphs and overlapping communities have all been investigated.
Few contributions however have encompassed all three types of graphs
simultaneously. In this paper we present a method that unifies community
detection for these three types of graphs while at the same time merges
partitioned and overlapping communities. Moreover, the results are
visualized in a way that allows for analysis and semantic interpretation.
For validation purposes this method is experimented on some well-known
simple benchmarks and then applied to real data:\textit{ photos and
tags in Facebook and Human Brain Tractography data. This last application
leads} to the possibility of applying community detection methods
to other fields such as data analysis with original enhanced performances.
\end{abstract}

\section{introduction}

Thanks to the growth of online social networks, community detection
has become an important field of research in computer sciences. Many
algorithms have been proposed (see several surveys on this topic in
\cite{Fortunato2009,Papadopoulos2011,Yang2010,Porter2009}). Most
of them take unipartite graphs as inputs and produce partitioned communities.
In unipartite graphs any node may share an edge with another node.
Other contributions have also explored bipartite graphs and directed
graphs. In bipartite graphs, nodes are separated in two sets and there
are only edges between nodes of different sets. In directed graphs
each link has a start node and an end node. These authors generally
introduce community detection methods which are specific for each
type of graphs, and sometimes for two types of graphs. In this paper
we present a method that encompasses all three types of graphs simultaneously
in a unique bipartite graph model. 

With this respect we consider Newman's modularity \cite{Newman2004}
and apply it to bipartite graphs. We show that this modularity model
can be directly applied to bipartite graphs with the side effect of
structurally linking objects of both node sets in the same communities.
This structural property is formally demonstrated in Annex 1.

In a second step this model is transformed into a unipartite graph
model. As a result any community detection algorithm for unipartite
graph may be applied. We chose for experiments the so-called Louvain
algorithm \cite{Blondel2008} which is known for its efficiency in
producing partitioned communities from extensive data sets. It is
also applicable to weighted and unweighted graphs. Our method extracts
communities where both types of nodes are associated. We show that
this result is semantically pertinent although it has been criticized by some
authors \cite{Murata2009,Guimera2007,Barber2007} who think that there 
should not be the same number of communities
in both sets. 
Moreover associating both types of nodes in the same
communities opens up new issues. It is possible to merge partitioned
and quantified overlapping communities in a unique view and then analyze
their structure with different perspectives. Indeed most community
detection algorithms such as Louvain use heuristics which lead to
local optima. With our approach we can identify and explain the final
organization and possibly correct some unwanted node assignments. 

In the following we use the term "`semantics"' for qualifying entities 
which are described by properties or attributes. Community detection is driven 
by properties that are shared between entities and consequently the resulting 
communities are semantically described by these properties. 

For validation and comparison with other authors the whole method
has been experimented on small traditional unipartite and bipartite
benchmarks. We have generated interesting insights which extend beyond
known results. We can then apply our method on real medium-sized bipartite
graphs, in a step that reveals significant properties such as overlapping
communities, community compactness and the role of inter-community
objects. These results are valuable when observed in data like people-photo
data sets targeted by our experiments. 

Beyond community detection, our method has also been applied to brain
data extracted through 'tractography' by a team of neurologists and
psycho-neurologists seeking to extract macro connections between different
brain areas. Our results were compared with those they obtained when
applying spectral clustering, a traditional data analysis method.
Although they were very similar, our method provided new insights
in the analysis. In conclusion we observe that after having borrowed
algorithms from data analysis methods, community detection may in
return offer new tools to these techniques.  We also successfully applied our method to 
most standard unipartite and bipartite graph benchmarks.

The next section will present a state-of-the-art on community detection
methods using different types of graphs. Section \ref{sec3} will
follow by focusing on a new method to unify all types of graphs; it
uses a definition of modularity for bipartite graphs directly derived
from modularity for unipartite graph which is presented in Annex 1
(section \ref{sec:Annex-1}). Section \ref{sec4} will then demonstrate
how our unifying method is particularly valuable in computing, visualizing
and analyzing partitioned and overlapping communities. Section \ref{sec5}
presents several practical results on different types of graph data
sets. The conclusion in section \ref{sec:Conclusion} discusses the
pros and cons of our method in the light of these experimental results.

\section{State of the art\label{sec:STATE-OF-THEart}}

\textit{\emph{As stated above, several state-of-the-art assessments
have already addressed the community detection problem: 
\cite{Papadopoulos2011,Porter2009,Yang2010,Fortunato2009}.
They are mainly focused on unipartite graph partitioning. The calculation
performed is based on maximizing a mathematical criterion, in most
cases modularity}} \cite{Newman2004}, representing the maximum number
of connections within each community and a minimum number of links
with external communities. Various methods have been developed to
identify the optimum, e.g. greedy algorithms \cite{Newman2004a,Noack2008},
spectral analysis \cite{Newman2006}, or a search for the most centric
edges \cite{Newman2004}. One of the most efficient greedy algorithm
for extracting partitioned communities from large (and possibly weighted)
graphs is Louvain \cite{Blondel2008}. In a very comprehensive \textit{\emph{state-of-the-art
report}} \cite{Fortunato2009} other new partitioned community detection
methods are described. 

The partitioning of communities, despite being mathematically attractive,
is not satisfactory to describe reality. Each individual has 'several
lives' and usually belongs to several communities based on family,
professional, and other activities. As such other methods more recently
take into account the possibility for overlapping communities. The
so-called k-clique percolation method \cite{Palla2005} detects overlapping
communities by allowing nodes to belong to multiple k-cliques. A more
recent method adapted to bipartite networks, and based on an extension
of the k-clique community detection algorithm is presented in \cite{SuneLehmannMartinSchwartzLarsKaiHansen2008}.
Several methods use local fitness optimization \cite{Lee2010c}\cite{Lancichinetti2009}.
The 'Label Propagation Algorithms' (LPA) are reported to be particularly
efficient \cite{Gregory2009}. \cite{Lee2010c} uses a greedy clique
expansion method to determine overlapping communities via a two-step
process: identify separated cliques and expand them for overlapping
by means of optimizing a local fitness criteria. \cite{Evans2009}
derives n order clique graphs from unipartite graphs to produce partitioned
and overlapping communities using Louvain algorithm. Some research
has provided results in the form of hypergraph communities such as
in\cite{Estrada2005,Chakraborty2012}. Other methods are found in
scientific papers, yet most of these are prone to major problems due
to computational complexity. More recently Wu \cite{Wu2012a} proposed
a fast overlapping community detection method for large real-world
unipartite networks. The method in \cite{Evans2009} presents some
common features with ours, albeit with a different strategy, since
it uses traditional partitioning algorithm to extract overlapping
communities. 

When considering semantics it becomes necessary to focus
 on bipartite or \textquotedblleft{}multi-partite\textquotedblright{}
graphs i.e. graphs whose nodes are divided into several subsets, and
whose edges only link nodes from different subsets. One example of
this type of graph is the set of photos from a Facebook account along
with their 'tags' \cite{MichelPlantie2011} or else the tripartite
network of epistemic graphs \cite{Roth2005} linking researchers,
their publications and keywords in these publications. Traditional
methods transform the multipartite graph into a unipartite graph by
assigning a link between two nodes should they share a common property.
In doing so however semantics is lost. Hence many researchers retain
the multiparty graph properties by extending the notion of modularity
to these types of graphs and then apply algorithms originally designed
for unipartite graphs \cite{Suzuki2009,NeubauerNicolas,Barber2007,Murata2009,Evans2009}\cite{LiuXin2010}.

\section{Unifying bipartite, directed and unipartite graphs\label{sec3}}

\subsection{Bipartite graphs partitioning \label{sub:Modularity-and-bimodularity}}

\subsubsection{Turning bipartite graphs into unipartite graphs.}

In formal terms, a bipartite graph $G=(U,V,E)$ is a graph $G'=(N,E)$
where node set $N$ is the union of two independent sets $U$ and
$V$ and moreover the edges only connect pairs of vertices $(u,v)$
where $u$ belongs to $U$ and $v$ belongs to $V$. 

$N=U\cup V$, $U\cap V=\emptyset$, $E\subseteq U\times V$.

$Let\: r=|U|\: and\: s=|V|,\: then\:|N|=n=r+s$

The unweighted biadjacency matrix of a bipartite graph $G=(U,V,E)$
is a $r\times s$ matrix $B$ in which $B_{i,j}=1\: iff(u_{i},v_{j})\in E\: and\: B_{i,j}=0\: iff(u_{i},v_{j})\notin E$. 

It must be pointed out that the row margins in $B$ represent the
degrees of nodes $u_{i}$ while the columns\textquoteright{} margins
represent the degrees of nodes $v_{j}$. Conversely, in $B^{t}$,
the transpose of $B$, row's margins represent the degrees of nodes
$v_{j}$ and columns\textquoteright{} margins represent the degrees
of nodes $u_{i}$. Let's now define the off-diagonal block square
matrix $A'$ :

$A'=\left(\begin{array}{cc}
0_{r} & B\\
B^{t} & 0_{s}
\end{array}\right)$ where $0_{r}$ is an all zero square matrix of order $r$ and $0_{s}$
is an all zero square matrix of order $s$. 

This symmetric matrix is the adjacency matrix of the unipartite graph
$\mathit{G'}$ where nodes' types are not distinguished. It is possible
to apply to $\mathit{G'}$ any algorithm for extracting communities
from unipartite graphs. $A'$ is also the off-diagonal adjacency matrix
of bipartite graph $\mathit{G}$. Consequently the communities which
are detected in $\mathit{G'}$ are also detected in $\mathit{G}$.
The question is to determine the validity of this side effect result:
what is the quality of partitioning for $\mathit{G}$ when applying
an unipartite graph partitioning algorithm on $\mathit{G}$'? Barber
\cite{Barber2007} and Liu/Murata \cite{LiuXin2010} have also introduced
the block matrix as a way of detecting communities in bipartite graphs.
However we see below that they do not take all consequences of this
approach.

\subsubsection{Extending modularity to bipartite graphs }

Modularity is an indicator often used to measure the quality of graph
partitions \cite{Newman2004}. First defined for unipartite graphs,
several modularity variants have been proposed for bipartite graph
partitioning and overlapping communities. More recently several authors
introduced modularity into bipartite graphs using a probabilistic analogy
with the modularity for unipartite graph which will be discussed below.
However when applying unipartite graph modularity optimization algorithms
to bipartite graphs, it is another expression of probabilistic modularity
presented hereafter. 

Let $G=(U,V,E)$ be a bipartite graph with its biadjacency matrix
$B$ and the unipartite graph $\mathit{G'}$ with the adjacency off-diagonal
block matrix $A'$. Let's consider Newman's modularity \cite{Newman2004}
for this graph $G'$. It is a function $Q$ of both matrix $A'$ and
the communities detected in $G'$ :

\begin{equation}
Q=\frac{1}{2m}\sum_{i,j}\left[A'_{ij}-\frac{k_{i}k_{j}}{2m}\right]\delta(c_{i},c_{j})\label{eq:1}
\end{equation}

where $A'_{ij}$ represents the weight of the edge between $i$ and
$j$, $k_{i}=\sum_{j}A'_{ij}$ is the sum of the weights of the edges
attached to vertex $i$, $c_{i}$ denotes the community to which vertex
$i$ is assigned, the Kronecker\textquoteright{}s function $\delta(u,v)$
equals $1$ if $u=v$ and $0$ otherwise and $m=1/2\sum_{ij}A'_{ij}$.
Hereafter we only consider binary graphs and weights are equal to
0 or 1.

After several transformations we show (see Annex 1, \prettyref{sec:Annex-1})
that this modularity can also be written using the biadjacency matrix
$B$ of the bipartite graph $G=(U,V,E)$:

\begin{equation}
Q^{B}=\frac{1}{m}\sum_{ij}[B_{ij}-\frac{(k_{i}+k_{j})\text{\texttwosuperior}}{4m}]\delta(c_{i},c_{j})\label{eq:9}
\end{equation}

where $k_{i}$ is the margin of row $i$ in $B$, $k_{j}$ the margin
of column $j$ in $B$ and $m=\sum_{ij}B_{ij}=\frac{1}{2}\times\sum_{ij}A'_{ij}=m$
in (\ref{eq:1}).

Another interesting formulation to be used is the following (Appendix
1, \prettyref{sec:Annex-1}): 

\begin{equation}
Q^{B}=\sum_{c}[\frac{|e_{c}|}{m}\text{\textendash}(\frac{(d_{u|c}+d_{v|c})}{2\times m})\text{\texttwosuperior}]\label{eq:3-2}
\end{equation}

where $|e_{c}|$ is the number of edges in community $c$, and $d_{w|c}$
is the degree of node $w$ belonging to $c$.

This formulation of modularity is the same as Newman's modularity
with more detailed information: it explicitly shows that both sets
of nodes are structurally associated in the same communities. 

Since in the general case $B$ is not symmetric, this definition thus
characterizes modularity for bipartite graphs after their extension
into unipartite graphs. It then becomes possible to apply any partitioning
algorithm for unipartite graphs to matrix $A'$ and obtain a result
where both types of nodes are bound in the same communities, except
in the case of singletons (i.e. nodes without edges). This definition
from unipartite graph modularity given that it is able to bind both
types of nodes, is compared in Section \ref{sub-Comparing-bimodularity-with}
with other authors' modularity models for bipartite graphs.

\subsubsection{Turning oriented graphs into bipartite graphs.}

A directed graph is of the form $G^{d}=(N,E^{d})$ where $N$ is a
set of nodes and $E^{d}$ is a set of ordered pairs of nodes belonging
to $N:\; E^{d}\subseteq N\times N$. From the model in \prettyref{eq:1}
Leicht \cite{Leicht2007} use probabilistic reasoning 'insights' to
derive the following modularity for directed network: 

\begin{equation}
Q=\frac{1}{m}\sum_{ij}\left[A_{ij}-\frac{k_{i}^{in}k_{j}^{out}}{m}\right]\delta(c_{i},c_{j})\label{eq:12}
\end{equation}
where $k_{i}^{in}$ and $k_{j}^{out}$ are the in - and out- degrees
of vertices $i$ and $j$, $A$ is the asymmetric adjacency matrix,
and $m=\sum_{ij}A_{ij}$ = $\sum_{i}k_{i}^{in}=\sum_{i}k_{j}^{out}$.
Symmetry is then restored and spectral optimization applied to extract
non-overlapping communities. This model leads to a node partition
that does not distinguish between the in and out roles; the nodes
are simply clustered within the various communities.

To compare these authors' method to ours, we transformed directed
graphs into bipartite graphs (this transformation was also suggested
in Guimera's work \cite{Guimera2007} when applying their method for
bipartite networks to directed graphs, as will be seen below). At
this point, let's differentiate the nodes' roles into $N\times N$.
Along these lines, we duplicate $N$ and consider two identical sets
$N^{out}$ and $N^{in}$. The original directed graph $G^{d}$ is
transformed into a bipartite graph $G=(N^{out},N^{in},E)$ in which
nodes appear twice depending on their 'out' or 'in' role and moreover
the asymmetric adjacency matrix $A$ plays the role of biadjacency
matrix $B$ in bipartite graphs. We can now define modularity for
directed graphs as follows:
\begin{equation}
Q^{B}=\frac{1}{m}\sum_{ij}[A_{ij}-\frac{(k_{i}^{in}+k_{j}^{out})\text{\texttwosuperior}}{4m}]\delta(c_{i},c_{j})\label{eq:5-1}
\end{equation}

After applying any algorithm for a unipartite graph on the corresponding
adjacency matrix $A'$ we obtain a partition where some nodes may
belong to the same community twice or instead may appear in two different
communities. Each model has its pros and cons. Leicht's model \cite{Leicht2007}
is preferable when seeking a single partition with no role distinction.
Our model is attractive when seeking to distinguish between 'in' and
'out' roles, e.g. between producers and customers where anyone can
play either role. The brain data example that follows will demonstrate
that our model is particularly well suited for analyzing real data.

\subsubsection{Turning unipartite graphs into bipartite graphs\label{sub:Bimodularity-for-unipartite graphs}}

In the above presentation, we introduced modularity for bipartite
graphs as a formal derivative of unipartite graph modularity. It is
dually possible to consider unipartite graphs as bipartite graphs,
and extract communities as if unipartite graphs were bipartite graphs.
To proceed, we must consider the original symmetric adjacency matrix
$A$ as an asymmetric biadjacency matrix $B$ (with the same nodes
on both dimensions) and build a new adjacency matrix $A'$ using the
original adjacency matrix $A$ twice on the off-diagonal, as if the
nodes had been cloned. When applying a unipartite graph partitioning
algorithm, we then obtain communities in which all nodes appear twice.
This method only works if we add to $A$ the unity matrix $I$ (with
the same dimensions as $A)$ before building $A'$. The first diagonal
in $A$ in fact only contains 0s since no loops are generally present
in a unipartite graph adjacency matrix. Semantically adding $I$ to
$A$ means that all objects will be linked to their respective clones
in $A'$. This is a necessary step in that when extracting communities,
the objects must drag their clones into the same communities in order
to maintain connectivity. In practice therefore, for unipartite graphs,
we build $A$' with $A+I$.

It may seem futile to perform such a transformation from a unipartite
graph to a bipartite one in order to find communities in unipartite
graphs given that for computing bipartite graph partitioning, we have
already made the extension into unipartite graphs using their (symmetric)
adjacency matrix. This transformation is nonetheless worthwhile for
several reasons. First, when appearing twice, nodes should be associated
with their clones. If the resulting communities do not display this
property, i.e. a node's clone lies in another community, then the
original matrix is not symmetric and can be considered as the adjacency
matrix of a directed graph. This conclusion has been applied to the
human brain tractography data clustering, which will be described
in the experimental section below.

Conversely, if we are sure that the original adjacency matrix is symmetric,
then a result where all nodes are associated with their clones in
the same communities would be a good indicator of the quality of the
clustering algorithm and moreover provides the opportunity to compare
our bipartite graph approach with other unipartite graph strategies.
This is also a method we introduced into our experiment (see the karate
and other applications below) for the purpose of verifying the validity of results.

Lastly, the most important benefit consists of building overlapping
communities and ownership functions for unipartite graphs using the
method explained in Section \ref{sec4} below. Although transforming
unipartite graphs into bipartite graphs requires more computation,
it also provides considerable information opening the way to semantic interpretation,
 which justifies its application in a variety of contexts.

\subsection{Comparison with other modularity models and partitioning algorithms
for bipartite graphs\label{sub-Comparing-bimodularity-with}}

Most modularity models which have been proposed in the literature
for bipartite graphs are inspired by Newman's modularity for unipartite
graphs. In some of them the objective is to distinguish the number
of communities in each type of nodes \cite{Guimera2007} \cite{MURATA2010}\cite{Suzuki2009}.
However there is a recent consensus on a probability null model introduced
by Barber \cite{Barber2007} which is very close to the original Newman's
modularity null model for unipartite graphs \cite{LiuXin2010}. Although
these authors introduce the same block matrix as we do, their modularity
model differs from ours.

After small transformations for unifying notation, Barber's model
( see \cite{Barber2009} equation 19) is the following: 

\begin{equation}
Q_{b}^{B}=\frac{1}{m}\sum_{i,j}[B_{ij}-\frac{k_{i}k_{j}}{m}]\delta(c_{i,}c_{j})\label{eq:6}
\end{equation}

This model is slightly different from our model: 

\begin{equation}
Q^{B}=\frac{1}{m}\sum_{ij}[B_{ij}-\frac{(k_{i}+k_{j})\text{\texttwosuperior}}{4m}]\delta(c_{i},c_{j})\label{eq:9-1}
\end{equation}

The formal difference is obvious and deserves some comments. Our modularity
expression is formally derived from Newman's unipartite modularity
model (see appendix).

As was shown in equation \ref{eq:3-2}, it is equivalent to considering
bipartite graphs as unpartite graphs with both types of graphs behaving
the same. 

We are therefore inclined to directly apply unipartite graph algorithms
which are based on this model and expect modularity optimization.
Conversely \cite{Barber2007}\cite{LiuXin2010}, although they consider
the same block matrix as we do, they specify a different null model which
is conceptually sound but not the result of a direct mathematical
derivation from the unipartite model. Therefore either these two definitions
are equivalent in terms of final optimization, or, if they are not,
Barber's model should be used with specific algorithms for bipartite
graphs, or with algorithms for unipartite graphs adapted to bipartite
graphs. 

If their interpretation is different, the effects of using either this
formula or the other can be observed according to two perspectives:
1) the number of communities in each set, 2) the node distribution
of each type in the communities. According to our definition of modularity,
both types of nodes are explicitly bound. Consequently when applying
any unipartite graph algorithm for detecting communities, both types
of nodes should have the same number of communities and, except for
singletons, they should be regrouped into the same communities (a
type of node should not be isolated in a community). This side effect
is not explicit in Equation \ref{eq:6}. However since in this equation
$\delta(c_{i},c_{j})$ specifies that the summation is applied to
both types of objects belonging to the same community, the side effect
is the same: optimizing the standard bipartite graph modularity should
yield a partitioning of both types of nodes in the same communities
(this analysis is also found in \cite{MURATA2010} : ``This definition
implicitly indicates that the numbers of communities of both types
are equal''). Both modularities should then produce the same results
in terms of node type distribution.

As far as the number of communities and node ownership are concerned,
it is more difficult to compare the results of both these models,
in particular if various algorithms are applied depending on the selected
model. For instance, in the Southern Women experiment described below,
we found 3 communities when applying Louvain, while Murata in \cite{LiuXin2010}
found four communities using their original LPAb+ algorithm. These
authors however only provided a quantitative evaluation via comparison
with other algorithms on computation performance and modularity optimization;
in contrast, we provide hereafter qualitative analysis as well, which
allows for semantics justification on the partitioning as will be showed in next section.

\section{Detection and analysis of community overlapping\label{sec4}}

\subsection{Adding semantics to communities}

The fact that both types of nodes are bound in their communities yields
several important results. First, in considering one type of nodes,
a community can be defined by associating a subset of nodes from the
other type. In other words, nodes from one set provides sense and
semantics for the grouping of nodes from the other set and moreover
may qualitatively explain regroupings, as will be seen below. This
semantic perspective has not been considered by any of the other authors,
a situation due to the fact that in other contributions, either the
number of communities differs for both types of nodes (e.g. \cite{Murata2009},
or else when both types of nodes contain the same number of communities
they are not bound in each community \cite{Guimera2007,Barber2007}. 

Binding both types of nodes into the same communities yields other
pertinent results. For one thing, it is possible to define belonging
functions and consequently obtain quantified overlapping communities.
In the following discussion, we will consider three possible belonging
functions, which may expose community overlapping in a different light.

\subsection{Probabilistic function}

Let's adopt the Southern Women's benchmark, which will be more thoroughly
described in Section \ref{sub:Southern-Women} below. Applying the
Louvain community detection algorithm for unipartite graphs yields
a partition where Women and Events are regrouped into three exclusive
communities. Let's call these communities $c_{1}$ , $c_{2}$ and
$c_{3}$. Now, let's suppose the fictitious case in which woman $w_{1}$
participated in events $e_{1}$, $e{}_{2}$, $e_{3}$ and $e_{4}$
. furthermore, $w_{1}$, $e_{1}$ and $e{}_{2}$ are classified in
$c_{1}$, while $e_{3}$ is classified in $c_{2}$ and $e_{4}$ is
classified in $c_{3}$. We can then define a probability function
as follows: 
\begin{equation}
P(u_{i}\in c)=\frac{1}{k_{i}}{\textstyle \sum_{j}B_{ij}\delta(c_{j})}
\end{equation}

where $c$ is a community, $k_{i}=\sum_{j}B_{ij}$ and $\delta(c_{j})=1$
if $v_{j}\in c$ or $\delta(c_{j})=0$ if $v_{j}\notin c$

In $P(u_{i}\in c)$ the numerator includes all edges linking $u_{i}$
to properties $v_{j}\in c$ and the denominator contains all edges
linking $u_{i}$ to all other nodes. With this function in the present
example the probability of $w_{1}$ being classified in community
$e_{1}$ equals $\frac{2}{4}$, and her probabilities of being classified
in $c_{2}$ and in $c_{3}$ are $\frac{1}{4}$ each. The probability
a node belongs to a given community is the percentage of its links
to this community as a proportion of the total number of links to
all communities. In other words, the greater the proportion of links
to a given community, the higher the expectation of belonging to this
community.

\subsection{Legitimacy function and overlapping communities\label{sub:Legitimacy-function}}

It is possible to add more meaning in order to decide which community
a given node should join. The legitimacy function serves to measure
the node involvement in a community and other results to show community overlapping.
The more strongly a node is linked to other nodes in a community,
the greater its legitimacy to belong to the particular community.
In the Southern Women's example, let's assume that after partitioning,
$c_{1}$ contains 7 events, $c_{2}$ 5 events and $c_{3}$ 2 events
(which is actually the case in the experiment presented below). Then,
$w_{1}$ would have a $\frac{2}{7}$ legitimacy for $c_{1}$, $\frac{1}{5}$
for $c_{2}$ and $\frac{1}{2}$ for $c_{3}$. The legitimacy function
can thus be formalized as follows:

\begin{equation}
L(u_{i}\in c)f=\frac{\sum{}_{j}B_{ij}\delta(c_{j})}{|\{v\in c\}|}
\end{equation}

where $c$ is a community, $\delta(c_{j})=1$ if $v_{j}\in c$ or
$\delta(c_{j})=0$ if $v_{j}\notin c$ 

The numerator in this expression is the same as the probabilistic
function numerator. Only the denominator is different.

\subsection{Reassignment Modularity function}

Reassigning node $w$ from $C_{1}$ to $C_{2}$ either increases or
decreases the modularity defined in Equation \prettyref{eq:9}. Such
a change is referred to as Reassignment Modularity ($RM_{w:C_{1}\rightarrow C_{2}}$).

The full development about this expression is exposed in Annex 2 (cf
section \ref{sec:Annex-2}). After simplification this expression
yields to:

\begin{equation}
RM_{w:C_{1}\rightarrow C_{2}}=\frac{1}{m}(l_{w|2}-l_{w|1})-\frac{1}{2m{}^{2}}[d_{w}^{2}+d_{w}(d_{C_{2}}-d_{C_{1}})]\label{eq:11}
\end{equation}
 
Reassignment is a very interesting measure. It allows detection of nodes that are not properly assigned to a community. Since most community detection algorithms are greedy algorithms some nodes may not be in a stable situation. The $RM$ value reveals  
unstable nodes and the community to which they should be assigned.

\section{Experimentation\label{sec5}}

This section will consider several benchmarks from various sources.
We begin by applying our method to two simple graphs: the so-called
\char`\"{}karate club\char`\"{} unipartite graph from \cite{Zachary1977}
shows friendship relations between members of a karate sport club;
and the \char`\"{}Southern Women\char`\"{} bipartite graph depicts
relations between southern American women participating in several
events. Our method is then applied to a medium-sized dataset extracted
from a real-world situation. For this purpose, we consider a bipartite
graph (people tagged on photos) drawn from a student's \char`\"{}Facebook\char`\"{}
account containing an average number of photos and people. Lastly,
this same method will be applied to human brain data in order to derive
dependencies between several areas in the brain. We also applied our method on several
well known unipartite and bipartite graph benchmarks as well as on big size benchmarks.

\subsection{Unipartite graph: Karate club}

The karate club graph \cite{Zachary1977} is a well-known benchmark
showing friendship relations between members of a karate club; it
is a unipartite graph on which many partitioning algorithms have been
experimented. Consequently, this set-up makes it possible not only
to verify that our method for bipartite graphs when applied to unipartite
graphs meets expected results, but also to assess the additional knowledge
extracted from overlapping.

\begin{figure}
\includegraphics[width=12cm]{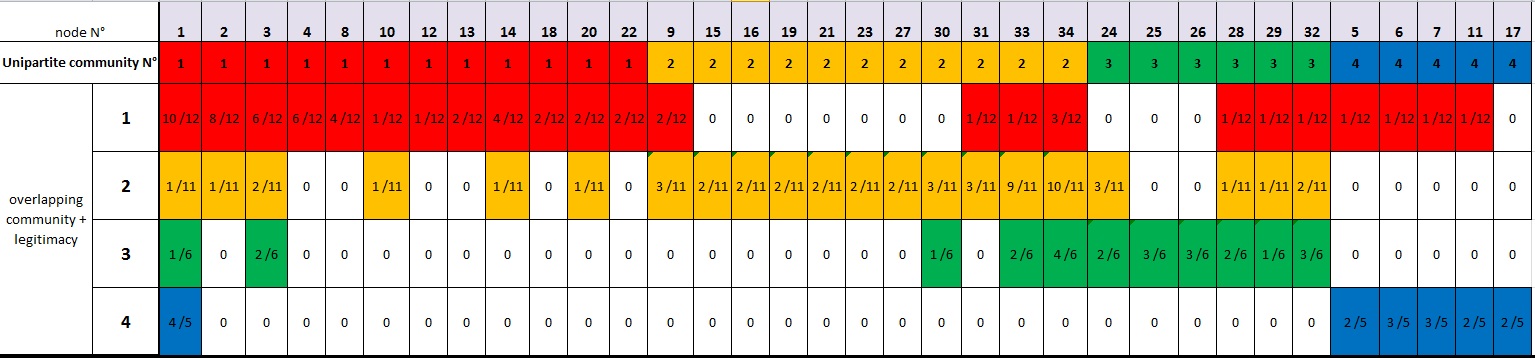}\caption{Karate club communities and legitimacy measures}
\label{karate}
\end{figure}

We began by directly applying the Louvain algorithm to the original
unipartite graph, represented by its adjacency matrix $A$. which
yielded four separate communities (as shown in \ref{karate}). These
are the same communities extracted by other authors, e.g. \cite{Newman2004}.
During a second experiment, we considered that the adjacency matrix
$A$ is in fact a biadjacency matrix $B$ which is representative
of a bipartite graph whose corresponding objects are the club members
and whose properties are also club members. An edge exists in the
bipartite graph between a club member-object and a club member-property
provided an edge is present between the two club members in the original
unipartite graph. The new $A'$ adjacency matrix is $A'=\left[\begin{array}{cc}
O_{r} & B\\
B^{t} & O_{s}
\end{array}\right]$, where $B=A+I$. and where $I$ is the identity matrix (as explained
in section \ref{sub:Bimodularity-for-unipartite graphs}). We
once again apply the Louvain algorithm to $A'$. 

\emph{Results. }As expected, these same four communities identified
in the unipartite graph have been extracted from the bipartite graph,
with the same individuals appearing twice in each community (see Figure
\ref{karate}). This initial result confirms the absence of bias when
transforming a unipartite graph into a bipartite one. The second result
is more pertinent because it reveals an overlap between communities
when considering legitimacy values. If we were to consider just the
cell colorings in the figure, an overlap would be observable whenever
at least one node from a community is linked to other nodes in another
community. The legitimacy values that indicate the involvement of
each node in each community offer an effective tool for identifying
and analyzing new features. Some slight differences have been noted
in works by other authors: for example, in page 2, Porter \cite{Porter2009}
placed node number 10 in the second community. In our case, this node
has been placed in the first community, though the legitimacy value
suggests that it should have been placed in the second one, in which
case the situation would be reversed in the second community and node
10 would have a legitimacy value that alters its placement in the
first community. Node 10 is thus in a hesitation mode between the
two communities.

To the best of our knowledge, this experiment represents the first
time Karate communities are shown as separate and overlapping. Partitioning
provides a practical way to observe communities; however, overlapping
reveals the extent to which partitioning reduces the amount of initial
information. With our method for example, it can be seen that some
nodes actually straddle several communities, e.g. node 10 in our experiment.

\begin{figure}
\includegraphics[width=5cm]{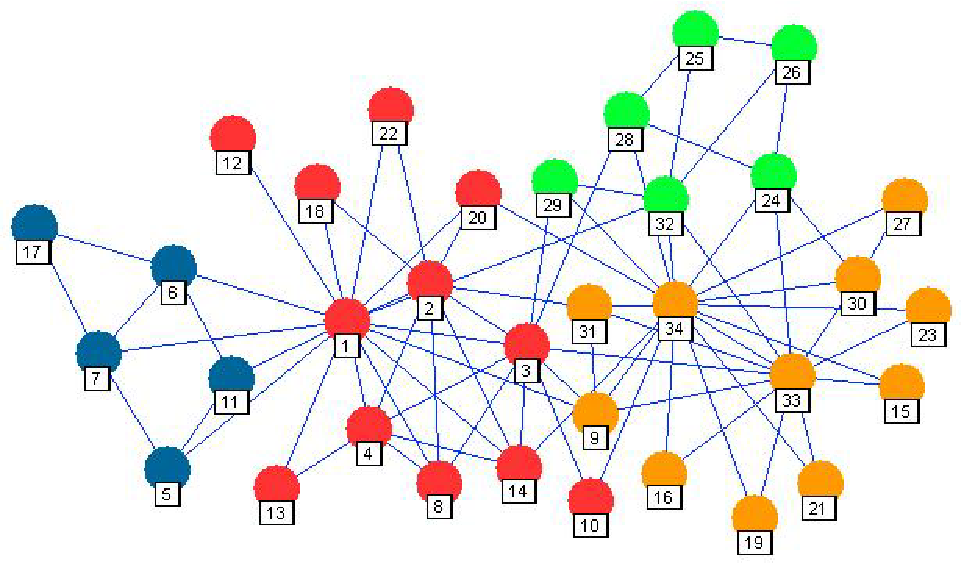}\caption{Karate club graph with partitioned communities}
\label{karate-1}
\end{figure}

\subsection{Unipartite graphs: other known benchmarks}

We have applied our method on several other well known unipartite
graphs, such as Dolphins and other benchmark graphs such as those
in \cite{Girvan2002}. As for the ``Karate'' case, we get the same
community partitions as Newman algorithm \cite{Newman2004}. \cite{Lancichinetti2008}
proposed a well known algorithm to generate benchmark graphs (also
used by \cite{Porter2009,Fortunato2009,Gregory2009} and others) where
communities are well identified. We used this algorithm to generate
30, 128, 500 and 1000 node such graphs to test our algorithms and
show the efficiency of our method. We do find the same number of communities
as Newman's algorithm since the modularity formula we use is directly
derived from Newman's one and we get the same analysis and results
as in \cite{Lancichinetti2008}. However we provide a very interesting
knowledge with supplementary data to observe node overlapping on these
communities.

The modularity has a limited resolution that depends on the number
of edges in the network \cite{Fortunato2007a}. We observed a main consequence
of the resolution limit: the modules in large networks may
have hidden substructures that require deeper investigations to reveal.

\subsection{Bipartite graph: Southern Women \label{sub:Southern-Women}}

This benchmark has been studied by most authors interested in checking
their partitioning algorithm for bipartite graphs. The goal here is
to partition, into various groups, 18 women who attended 14 social
events according to their level of participation in these events.
In his well-known cross-sectional study, \cite{Freeman} compared
results from 21 authors, most of whom identified two groups.

\begin{figure}
\includegraphics[width=10cm]{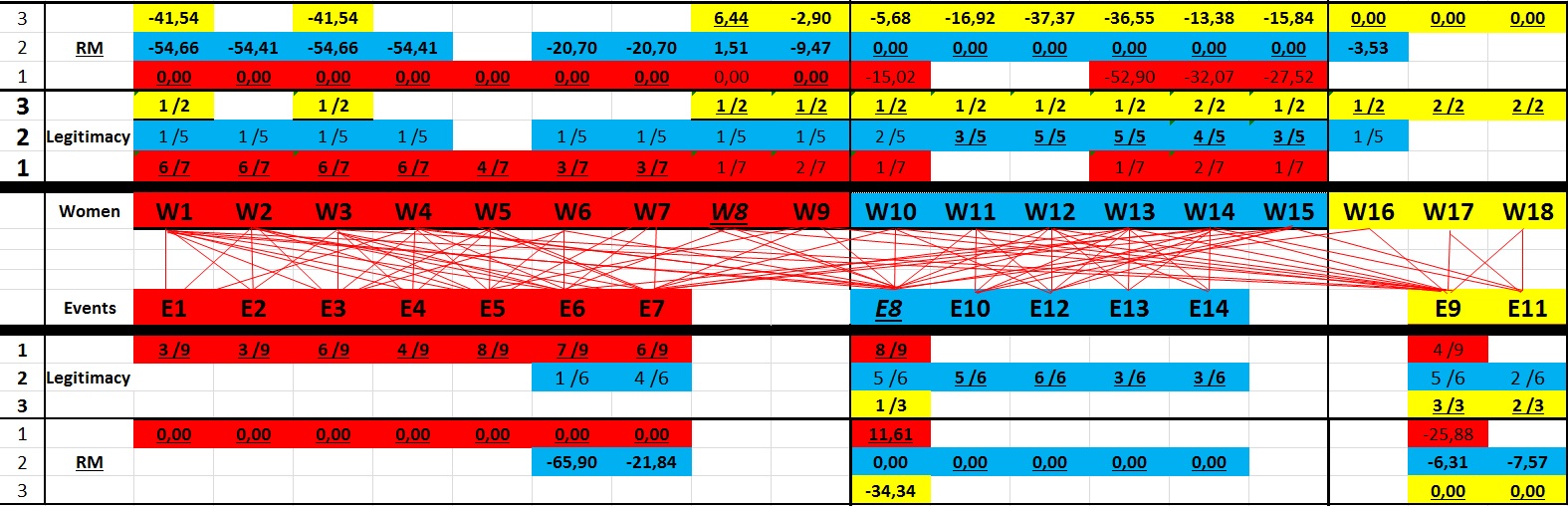}\caption{Women Events communities with legitimacy and Reassignment modularity
measures}
\label{WE}
\end{figure}

\emph{Results. }In Figure \ref{WE}, the bipartite graph is depicted
as a bi-layer graph in the middle with women at the top and events
below; moreover, the edges between women and events represent woman-event
participations. Three clusters with associated women and events have
been found and eventually shown with red, blue and yellow colorings.
This result is more accurate than the majority of results presented
in \cite{Freeman}; only one author found three female communities.
Beyond mere partitioning, Figure \ref{WE} presents overlapping communities
using two overlapping functions, namely legitimacy and reassignment
modularity (RM). Legitimacy and RM for women are placed just above
female partitioning; for events, both are symmetrically shown below
event partitioning. As expected, reassignment in the same community
produces a zero RM value. The best values for legitimacy and RM have
been underscored. Only the values of woman 8 and event 8 indicate
that they could have been in another community. This is the outcome
of early assignment during the first Louvain phase for entities with
equal or nearly equal probabilities across several communities. It
can be observed in \cite{Freeman} that woman 8's community is also
debated by several authors; our results appear to be particularly
pertinent in terms of both partitioning and overlapping.

The fact that women and events are correlated may be considered to
cause a bias, such as in the number of communities. When comparing
our results to those of other authors however, the merging of our
blue and yellow communities produces their corresponding second community.
In their trial designed to obtain a varying number of communities
in both sets, Suzuki \cite{Suzuki2009} found a large number of singletons.
Their results were far from those presented in \cite{Freeman}, while
ours were compatible and more highly detailed.

In conclusion, results on the Southern Women's benchmark are particularly
relevant. Moreover, our visualization enables observing community
partitioning, overlapping and possible assignment contradictions.
The application of reassignment for better modularity optimization
will be tested in a subsequent work.

\subsection{Bipartite graph: Facebook account}

\begin{figure}
\includegraphics[height=3cm]{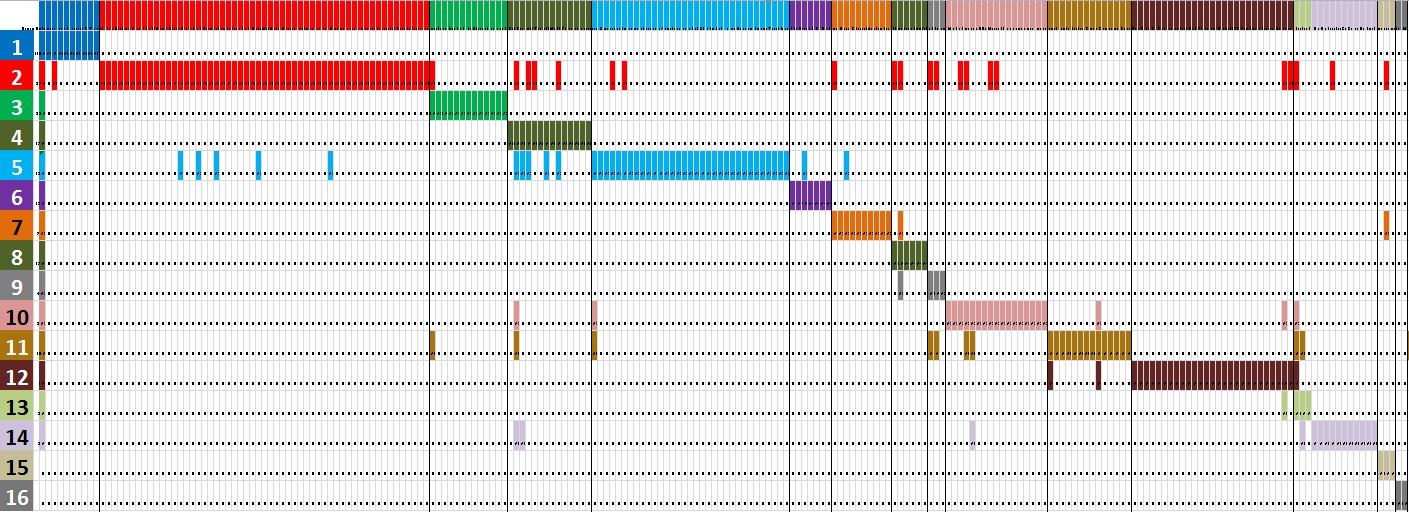}

\caption{Facebook account communities with overlapping}
\label{facebook}
\end{figure}

In a Facebook (FB) account, several types of informations may be extracted.
We extracted and evaluated only data coming from FB photo albums with
its tags. We did not use friendship relations. Three Facebook photo
files were downloaded from various Facebook (FB) accounts. All these
files were extracted with the consent of their owners, none of whom
were members of the research team. A person was considered to be linked
to a photo if he/she had been tagged in the photo. We then have a
bipartite graph composed of two type of nodes : persons and photos.
Community extraction using our method reveals some common features
among the datasets. These features are shown in Figure \textit{\emph{\ref{facebook}
for one FB photo file, in which 274 people could be identified in
a total of 644 photos.}}

\emph{Results. }Communities are seldom overlapping, which supports
the notion that the photos were taken at different times in the owner's
life (this is to be confirmed in a forthcoming study). When the owner
was asked to comment on the communities, two main observations were
submitted. The various groups of people were indeed consistent, yet
with one exception. The owner was associated in the partition with
a group she had met on only a few occasions and not associated with
other groups of close friends. An analysis of the results provided
a good explanation, which is partially displayed in Figure \ref{facebook}.
From this view, the FB account owner is in the first community on
the left, yet she is also present in most of the other communities
(see grey color levels in the first column). Although at first glance
it might be assumed that she is not part of other communities, our
visualization indicates that such is not the case. She is present
in most communities, even though she is mainly identified in the first
one. Three types of photos can be distinguished in this first community.
More than 200 photos only contain the owner's tag, plus a few photos
with unique tags of another community member; for every other person,
at least one photo tags him/her with the owner. This first community
has in fact been built from the first group with photos of unique
owner's tags associated with the owner. The owner's tag thus encompasses
photos containing two people, one of whom is the owner. It turns out
that this group is predominantly the owner's group.

In conclusion, partitioning only the bipartite graph would have produced
a major pitfall: the owner would have been isolated in a community
that is not his/her top preference. With our method, merging partitioning
and overlapping exposes better multiple regroupings with broader affinities.
Other communities also showed high consistency when considering the
photos: each community was associated with some particular event responsible
for gathering a group of the FB account owner's friends.

\subsection{Bipartite graph: Brain Data }

Our method was initially designed for human community detection and
analysis. In this experiment, we have demonstrated how it can be applied
to other data analysis techniques as well. The brain dataset was collected
on a single patient by a research team affiliated with the \char`\"{}Human
Connectome\char`\"{} project working on brain tractography techniques
\cite{Catani2012}. These techniques use Magnetic Resonance Imaging
(MRI) and Diffusion Tensor Imaging (DTI) to explore white matter tracks
between brain regions. Probabilistic tractography produces 'connectivity'
matrices between Regions Of Interest (ROI) in the brain. For the case
we studied, 'seed' ROIs were located in the occipital lobe and 'target'
ROIs throughout the entire brain. The goal here was to detect possible
brain areas in the occipital lobe through ROI clustering on the basis
of similar track behavior. In \cite{Catani2012}, the research team
used Spectral Clustering (SC) to combine ROIs. It is interesting to
note that SC is one of numerous techniques that have traditionally
been applied in social community detection, e.g. by Bonacich on the
Southern Women's benchmark \cite{Freeman}. SC results are limited
to community partitioning (though in theory overlapping could also
be computed). The goal was to experiment with our method and produce
both partitioning and overlapping analyzes of brain areas.

\begin{figure}
\includegraphics[width=12cm]{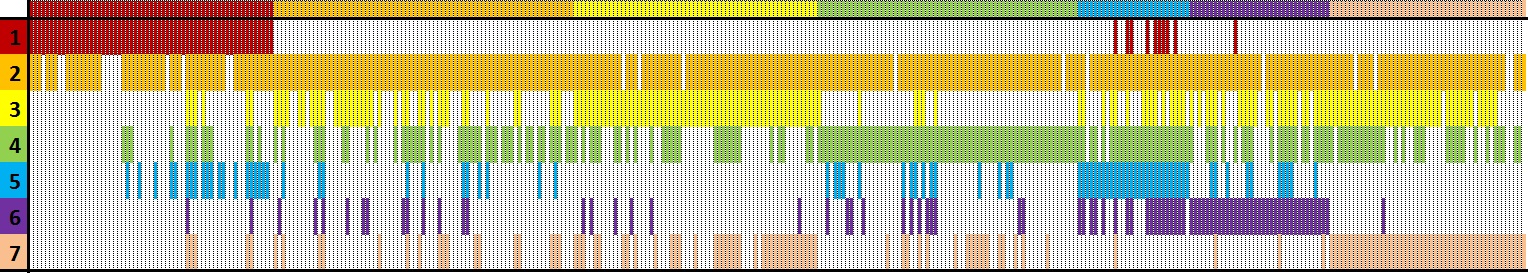}\caption{Brain data communities with overlapping}
\label{brain}
\end{figure}

The original matrix contained 1,914 rows and 374 columns, with cells
denoting the probabilities of linkage between ROIs. We considered
this matrix as a bipartite graph biadjacency matrix with weighted
values and then applied our community detection method. Figure \ref{brain}
presents the results of ROI community partitioning and overlapping.
Each color in the first row is associated with a community that gathers
several ROIs. Each ROI is represented by a column that indicates its
belonging to the other communities. When a cell is highlighted with
a color, a nonzero overlapping value exists for both this ROI and
the corresponding community (with community numbers being plotted
on the left-hand side of the figure). This value has been computed
with the legitimacy function, which has been extended to the weighted
edges, i.e. the weighted sum of values from cerebral hemisphere zones
(ELF) within the selected community. Each community is associated
with a threshold value corresponding to the maximum weighted legitimacy
above which the community would lose a full member. For each community,
this threshold value is automatically computed in order to include
all ROI members of the community.

\emph{Results. }We found 7 communities when neurologists selected
8 clusters with SC and after choosing the most significant eigenvectors
on a scree test. Let's observe that two communities overlap heavily
on all others, which thus overlap to a lesser extent. Figure \ref{brain}
confirms the strong interest in this set-up that simultaneously exhibits
overlapping and non-overlapping data. These results have been taken
into account by a team of neurological researchers as different observations
recorded on brain parcellation.

\subsection{Bipartite graphs: others known benchmarks}

Bipartite graph datasets are not easy to find in litterature. We also
tested our algorithms with bipartite networks used as benchmark networks
in \cite{Barber2007}. One of them is the network benchmark describing
corporate interlocks in Scotland in the early twentieth century. The
data set characterizes 108 Scottish firms during 1904-5, detailing
the corporate sector, capital, and board of directors for each firm.
The data set includes only those board members who held multiple directorships,
totaling 136 individuals. Barber found ``roughly'' (sic) 20 communities,
whereas we find 15 communities and provide very interesting knowledge
about overlapping for these communities. We obtained a global modularity
of 0.71038 whereas Barber found a smaller value of 0.56634.

To evaluate scalability on our method we tested a rather big co-authorship
bipartite dataset to detect scientific communities extracted from
the well known PubMed (http://www.ncbi.nlm.nih.gov/pubmed) biomedical
scientific litterature online library. Our dataset was composed of
30,000 persons and more than 80,000 scientific papers. We extracted
184 communities of average 670 members in about 3 seconds, with interesting
overlapping information. Regarding resolution limit mentionned earlier,
the modularity method applied to bipartite graphs has a similar limit,
with similar consequences.

\section{Discussion and new perpectives}

The above experiments show that our method is able to find overlapping
communities in different types of graphs. Moreover, it is able to
measure the degree of membership for each node to each community.
We then get a first semantic interpretation of each node in terms
of community membership. These results are obtained through the use
of the off-diagonal block square matrix $A^{'}$. Several other methods
may compute modularity by using directly a graph structure without
building any off-diagonal block matrix. For example LPA based methods
\cite{Gregory2009,LiuXin2010} which use Barber's modularity definition
for bipartite graphs may work directly with the graph structure. However
the results that are presented in these papers are different. They
find 4 communities for the Women Events dataset instead of 3 in our
case. Since Barber's modularity expression is different from ours,
it is difficult to compare these different results.

Louvain algorithm which uses Newman's modularity formula is adapted
to monopartite graphs. Since the approach with the block matrix requires
more data and computation, we also tried applying Louvain algorithm
directly on to the biadjacency matrix $B$. We tested it on the same
two bipartite data sets: Women Events and Facebook. Surprisingly,
we got the same results as those with the block square matrix $A^{'}$.
This experiment suggests the possibility of directly applying unipartite
graph models onto bipartite graph models with unipartite graph modularity.
Moreover our method with the block matrix $A^{'}$ could be a good
means for validating this possibility. 

This counter intuitive conclusion needs more experiments and more
theoretical proof. Particularly since other authors use Barber's model
which is specifically adapted for bipartite graphs. Future work will
deeper investigate this possibility of directly applying unipartite
graph methods to bipartite graphs.

\section{Conclusion\label{sec:Conclusion}}

In this paper, we have demonstrated the feasibility of unifying bipartite
graphs, directed graphs and unipartite graphs under a common unipartite graph model. 
It was then proved that any unipartite graph partitioning algorithm
aiming at optimizing the standard unipartite modularity model leads 
to a bipartite graph partitioning, wherein both types of nodes
are bound in the communities. In the special case of directed graphs,
nodes appear twice in potentially different communities depending
on their roles; for unipartite graphs, nodes are cloned and appear
with their clones in the same communities.

We also introduced the possibility of unifying in a single view, 
the partitioning and the overlapping communities. This development
is possible thanks to associating both types of nodes in the communities.
Moreover, overlapping can be characterized through several functions
presenting different interpretations. For instance, it is possible
to identify those nodes that define the community cores, i.e. those
who belong exclusively to just one community and, conversely, those
who serve as bridges between different communities. We also introduced 
reassignment values which open up the possibility of improving partitioning results. 
Practically speaking, when applying our method to various benchmarks
and datasets, we are able to extract meaningful communities and display
surprising overlapping properties when other authors limit their goal to identifying communities. 
We extend far beyond this point and
provide tools for analyzing and interpreting results. 

Lastly, we introduced an essential result after experimenting on real
brain datasets, supplied by a research team from the Connectome project.
Historically authors dealing with community detection problems used to borrow 
their methods from data or graph analysis such as hierarchical clustering, 
clique enumeration or spectral analysis. Recent community detection approaches 
based on modularity optimization use original methods (Louvain, label propagation). 
We showed that these methods could also be applied to data analysis with good results. 
Moreover these results can be obtained without the need to choose parameters such as 
the number of clusters, or a threshold value. It is of particular interest to note that 
after borrowing their methods from other scientific domains, community detection techniques 
are now enough mature for providing these domains with new original performing methods.

In the future we will continue exploring cross fertilization between community detection techniques 
and other scientific  domains. In particular we will use Nash equilibrium for studying community 
stability through the reassignment value we introduced in this paper. Indeed we think that community 
stability could be another quality criteria along with  modularity optimization for driving 
and assessing community detection algorithms' performances.

\subsection*{Acknowledgments}

The authors would like to thank the Connectome research team, as part
of the ``CAFO'' project (ANR-09-RPDOC-004-01 project) as well as
the CRICM UPMC U975/UMRS 975/UMR 7225 research group for providing
the original brain dataset.

\bibliographystyle{plain}
\bibliography{communities}

\begin{thebibliography}{10}

\bibitem{Barber2007}
Michael Barber.
\newblock {Modularity and community detection in bipartite networks}.
\newblock {\em Physical Review E}, 76(6):1--9, 2007.

\bibitem{Barber2009}
Michael~J Barber and John~W Clark.
\newblock {Detecting network communities by propagating labels under
  constraints}.
\newblock {\em Physical Review E - Statistical, Nonlinear and Soft Matter
  Physics}, 80(2 Pt 2):026129, 2009.

\bibitem{Blondel2008}
Vincent~D Blondel, Jean-Loup Guillaume, Renaud Lambiotte, and Etienne Lefebvre.
\newblock {Fast unfolding of communities in large networks}.
\newblock {\em Journal of Statistical Mechanics: Theory and Experiment},
  2008(10):P10008, October 2008.

\bibitem{Catani2012}
Marco Catani and Michel~Thiebaut de~Schotten.
\newblock {\em {Atlas of human brain connections}}.
\newblock Oxford University Press, 2012, 2012.

\bibitem{Chakraborty2012}
Abhijnan Chakraborty, Saptarshi Ghosh, and Niloy Ganguly.
\newblock {Detecting overlapping communities in folksonomies}.
\newblock In {\em Proceedings of the 23rd ACM conference on Hypertext and
  social media HT 12}, page 213. ACM Press, 2012.

\bibitem{Estrada2005}
Ernesto Estrada and Juan~A Rodriguez-Velazquez.
\newblock {Complex Networks as Hypergraphs}.
\newblock {\em Systems Research}, page~16, 2005.

\bibitem{Evans2009}
T~S Evans and R~Lambiotte.
\newblock {Line Graphs, Link Partitions and Overlapping Communities}.
\newblock {\em Physical Review E}, 80(1):9, 2009.

\bibitem{Fortunato2009}
Santo Fortunato.
\newblock {Community detection in graphs}.
\newblock {\em Physics Reports}, 486(3-5):103, June 2009.

\bibitem{Fortunato2007a}
Santo Fortunato and Marc Barth\'{e}lemy.
\newblock {Resolution limit in community detection.}
\newblock {\em Proceedings of the National Academy of Sciences of the United
  States of America}, 104(1):36--41, 2007.

\bibitem{Freeman}
Linton~C. Freeman.
\newblock {Finding social groups: A meta-analysis of the southern women data}.
\newblock In {\em Dynamic Social Network Modeling and Analysis. The National
  Academies}, pages 39----97. Press, 2003.

\bibitem{Girvan2002}
M.~Girvan and M~E~J Newman.
\newblock {Community structure in social and biological networks}.
\newblock {\em Proceedings of the National Academy of Sciences of the United
  States of America}, 99(12):7821--7826, 2002.

\bibitem{Gregory2009}
Steve Gregory.
\newblock {Finding overlapping communities in networks by label propagation}.
\newblock {\em New Journal of Physics}, 12(10):103018, 2009.

\bibitem{Guimera2007}
Roger Guimer\`{a}, Marta Sales-Pardo, and Lu\'{\i}s Amaral.
\newblock {Module identification in bipartite and directed networks}.
\newblock {\em Physical Review E}, 76(3), September 2007.

\bibitem{Lancichinetti2009}
Andrea Lancichinetti, Santo Fortunato, and J\'{a}nos Kert\'{e}sz.
\newblock {Detecting the overlapping and hierarchical community structure in
  complex networks}.
\newblock {\em New Journal of Physics}, 11(3):033015, March 2009.

\bibitem{Lancichinetti2008}
Andrea Lancichinetti, Santo Fortunato, and Filippo Radicchi.
\newblock {Benchmark graphs for testing community detection algorithms.}
\newblock {\em Physical Review E - Statistical, Nonlinear and Soft Matter
  Physics}, 78(4 Pt 2):6, 2008.

\bibitem{Lee2010c}
Conrad Lee, Fergal Reid, Aaron McDaid, and Neil Hurley.
\newblock {Detecting highly overlapping community structure by greedy clique
  expansion}.
\newblock {\em 4th Workshop on Social Network Mining and Analysis SNAKDD10},
  10:10, 2010.

\bibitem{Leicht2007}
E~A Leicht and M~E~J Newman.
\newblock {Community structure in directed networks}.
\newblock {\em Physical Review Letters}, 100(11):118703, 2007.

\bibitem{LiuXin2010}
{Liu Xin} and {Murata Tsuyoshi}.
\newblock {An Efficient Algorithm for Optimizing Bipartite Modularity in
  Bipartite Networks}.
\newblock {\em Journal of Advanced Computational Intelligence and Intelligent
  Informatics}, 14(4):408--415, 2010.

\bibitem{MichelPlantie2011}
{Michel Planti\'{e}} and {Michel Crampes}.
\newblock {Mining social networks and their visual semantics from social
  photos}.
\newblock {\em International Journal of Computer science \& Applications},
  VIII(II):102--117, 2011.

\bibitem{Murata2009}
Tsuyoshi Murata.
\newblock {Modularities for bipartite networks}.
\newblock {\em Proceedings of the 20th ACM conference on Hypertext and
  hypermedia HT 09}, 90(6):245--250, 2009.

\bibitem{MURATA2010}
Tsuyoshi Murata.
\newblock {Detecting communities from tripartite networks}.
\newblock {\em WWW}, pages 0--1, 2010.

\bibitem{NeubauerNicolas}
{Neubauer Nicolas} and {Obermayer Klaus}.
\newblock {Towards Community Detection in k-Partite k-Uniform Hypergraphs}.
\newblock In {\em Proceedings NIPS 2009 \ldots}, 2009.

\bibitem{Newman2004a}
Mark Newman.
\newblock {Fast algorithm for detecting community structure in networks}.
\newblock {\em Physical Review E}, 69(6), June 2004.

\bibitem{Newman2006}
Mark Newman.
\newblock {Finding community structure in networks using the eigenvectors of
  matrices}.
\newblock {\em Physical Review E - Statistical, Nonlinear and Soft Matter
  Physics}, 74(3 Pt 2):036104, 2006.

\bibitem{Newman2004}
Mark Newman and M.~Girvan.
\newblock {Finding and evaluating community structure in networks}.
\newblock {\em Physical Review E}, 69(2), February 2004.

\bibitem{Noack2008}
Andreas Noack and Randolf Rotta.
\newblock {Multi-level algorithms for modularity clustering}.
\newblock {\em arXiv}, page~12, December 2008.

\bibitem{Palla2005}
Gergely Palla, Imre Der\'{e}nyi, Ill\'{e}s Farkas, and Tam\'{a}s Vicsek.
\newblock {Uncovering the overlapping community structure of complex networks
  in nature and society.}
\newblock {\em Nature}, 435(7043):814--8, June 2005.

\bibitem{Papadopoulos2011}
S~Papadopoulos, Y~Kompatsiaris, A~Vakali, and P~Spyridonos.
\newblock {Community detection in Social Media}.
\newblock {\em Data Mining and Knowledge Discovery}, 1(June):1--40, 2011.

\bibitem{Porter2009}
Mason~A. Porter, Jukka-Pekka Onnela, and Peter~J. Mucha.
\newblock {Communities in Networks}, 2009.

\bibitem{Roth2005}
Camille Roth and Paul Bourgine.
\newblock {Epistemic Communities: Description and Hierarchic Categorization}.
\newblock {\em Mathematical Population Studies: An International Journal of
  Mathematical Demography}, 12(2):107--130, 2005.

\bibitem{SuneLehmannMartinSchwartzLarsKaiHansen2008}
{Sune Lehmann,Martin Schwartz,Lars Kai Hansen}.
\newblock {Biclique communities.}
\newblock {\em Physical review. E, Statistical, nonlinear, and soft matter
  physics}, 78(1 Pt 2), 2008.

\bibitem{Suzuki2009}
Kenta Suzuki and Ken Wakita.
\newblock {Extracting Multi-facet Community Structure from Bipartite Networks}.
\newblock {\em 2009 International Conference on Computational Science and
  Engineering}, 4:312--319, 2009.

\bibitem{Wu2012a}
Zhihao Wu, Youfang Lin, Huaiyu Wan, Shengfeng Tian, and Keyun Hu.
\newblock {Efficient overlapping community detection in huge real-world
  networks}.
\newblock {\em Physica A: Statistical Mechanics and its Applications},
  391(7):2475 -- 2490, 2012.

\bibitem{Yang2010}
Bo~Yang, Dayou Liu, Jiming Liu, and Borko Furht.
\newblock {\em {Discovering communities from Social Networks: Methodologies and
  Applications }}.
\newblock Springer US, Boston, MA, 2010.

\bibitem{Zachary1977}
W~W Zachary.
\newblock {An information flow model for conflict and fission in small groups}.
\newblock {\em Journal of Anthropological Research}, 33(4):452--473, 1977.

\end{thebibliography}

\section{Annex 1\label{sec:Annex-1}}

\subsection{New use of Newman modularity \label{sub:From-modularity-to} }

In this Annex, we will provide full details of the demonstration
that yielded Equation \prettyref{eq:9}.

For the sake of convenience, let's use the definition of unipartite
graph modularity offered in Newman \cite{Leicht2007}. It is a function
$Q$ of matrix $A^{'}$ and the communities detected in $G$ \cite{Newman2004}:

\begin{equation}
Q=\frac{1}{2m}\sum_{i,j}\left[A_{ij}^{'}-\frac{k_{i}k_{j}}{2m}\right]\delta(c_{i},c_{j})\label{eq:3-1}
\end{equation}

where $A_{ij}^{'}$ denotes the weight of the edge between $i$ and
$j$, $k_{i}=\sum_{j}A_{ij}^{'}$ is the sum of the weights of edges
attached to vertex $i$, $c_{i}$ is the community to which vertex
$i$ has been assigned, the Kronecker\textquoteright{}s function $\delta(u,v)$
equals $1$ if $u=v$ and $0$ otherwise and $m=1/2\sum_{ij}A_{ij}^{'}$.

In our particular case (i.e. where $A^{'}$ is the off-diagonal block
adjacency matrix of a bipartite graph), we apply the following transformations:

Let's rename $i_{1}$ as index $i$ when $1\leq i\leq r$ and $i_{2}$
when $r<i\leq r+s$. Conversely, let's rename $j_{1}$ the index $j$
when $1\leq j\leq r$ and $j_{2}$ when $r<j\leq r+s$.

To avoid confusion between the $A^{'}$'s indices and $B$\textquoteright{}s
indices let's rename $B$ indices $i_{b}$ and $j_{b}$ : $1\leq i_{b}\leq r$
and $1\leq j_{b}\leq s$ (see a representation of $A$ matrix below
(Figure \ref{eq:4-1}))

\begin{equation}
A'=\begin{array}{c|c|c|c|c}
A^{'}\: indexes\downarrow\rightarrow & ....j_{1}.... & ....j_{2}.... & \\
\hline ... &  &  & ...\\
i_{1} & O_{r} & B & i_{b} & r\: rows\\
... &  &  & ...\\
\hline ... &  &  & ...\\
i_{2} & B^{t} & O_{s} & j_{b} & s\: rows\\
... &  &  & ...\\
\hline  & ....i{}_{b}.... & ....j_{b}.... & \leftarrow\uparrow B\: indexes\\
\hline  & r\: columns & s\: columns & 
\end{array}\label{eq:4-1}
\end{equation}

Let's call $k_{i_{b}}$ the margin of row $i_{b}$ in $B$ and $k_{j_{b}}$
the margin of column $j_{b}$ in $B$.

\begin{equation}
k_{i_{b}}=\sum_{j_{b}}B_{i_{b}j_{b}}=\sum_{j_{2}}A_{i_{1}j_{2}}^{'}=\sum_{i_{2}}A_{i_{2}j_{1}}^{'},\: where\: i_{b}=i_{1}=j_{1}\label{eq:2a}
\end{equation}

\begin{equation}
k_{j_{b}}=\sum_{i_{b}}B_{i_{b}j_{b}}=\sum_{i_{1}}A_{i_{1}j_{2}}^{'}=\sum_{j_{1}}A_{i_{2}j_{1}}^{'},\: where\: j_{b}=i_{2}\text{\textendash}r=j_{2}\text{\textendash}r\label{eq:2b}
\end{equation}

$k_{i_{b}}$is the degree of node $u_{i_{b}}$, $k_{j_{b}}$ is the
degree of node $v_{j_{b}}$. Let's define $k_{i/j_{1}}=\sum{}_{j_{1}}A_{ij_{1}}^{'}\: and\: k_{i/j_{2}}=\sum_{j_{2}}A_{ij_{2}}^{'}$.
Conversely : $k_{j/i_{1}}=\sum_{i_{1}}A_{ji_{1}}^{'}\: and\: k_{j/i_{2}}=\sum_{i_{2}}A_{ji_{2}}^{'}$.
Hence : $k_{i}=\sum_{j}A_{ij}^{'}=k_{i/j_{1}}+k_{i/j_{2}}$, $k_{j}=\sum_{i}A_{ij}^{'}=k_{j/i_{1}}+k_{j/i_{2}}$.

By taking into account the structure and properties of $A'$ in \prettyref{eq:2a}
and \prettyref{eq:2b} for the indices we derive the following properties
\label{eq:3}: 

$k_{i/j_{1}}$ has non-zero values only for $i=i_{2}$, with $k_{j_{b}}$
the degree of node $v_{j_{b}}$:

\begin{equation}
k_{i/j_{1}}=k_{i_{2}/j_{1}}=\sum_{j_{1}}A_{i_{2}j_{1}}^{'}=\sum_{i_{1}}A_{i_{1}j_{2}}^{'}=k_{j_{2}/i_{1}}=k_{j_{b}}\label{eq:2b-1}
\end{equation}

$k_{i/j_{2}}$ has non-zero values only for $i=i_{1},$ with $k_{i{}_{b}}$
the degree of node $u_{i{}_{b}}$:

\begin{equation}
k_{i/j_{2}}=k_{i_{1}/j_{2}}=\sum_{j_{2}}A_{i_{1}j_{2}}^{'}=\sum_{i_{2}}A_{i_{2}j_{1}}^{'}=k_{j_{1}/i_{2}}=k_{i_{b}}\label{eq:2b-1-1}
\end{equation}

Moreover and more directly: $k_{j/i_{1}}$ offers values only for
$j=j_{2}$: $k_{j/i_{1}}=k_{j_{2}/i_{1}}=k_{i_{2}/j_{1}}=k_{j_{b}}$,
the degree of node $v_{j_{b}}$. $k_{j/i_{2}}$ offers values only
for $j=j_{1}$: $k_{j/i_{2}}=k_{j_{1}/i_{2}}=k_{i_{1}/j_{2}}=k_{i_{b}}$,
the degree of node $u_{i_{b}}$.

\subsection{Analyzing second part of $Q$ in equation \prettyref{eq:3-1}}

Using these properties of matrix $A^{'}$, it is now possible to analyze
$\sum_{ij}k_{i}k_{j}$. in equation \prettyref{eq:3-1}.

Next, by developing $k_{i}$ and $k_{j}$ in $A^{'}$ we obtain: $\sum_{ij}k_{i}k_{j}=\sum_{ij}(k_{i/j_{1}}+k_{i/j_{2}})(k_{j/i_{1}}+k_{j/i_{2}})$

$=\sum_{ij}k_{i/j_{1}}k_{j/i_{1}}+\sum_{ij}k_{i/j_{2}}k_{j/i_{2}}+\sum_{ij}k_{i/j_{1}}k_{j/i_{2}}+\sum_{ij}k_{i/j_{2}}k_{j/i_{1}}$

\begin{equation}
=\sum_{i_{2}j_{2}}k_{i_{2}/j_{1}}k_{j_{2}/i_{1}}+\sum_{i_{1}j_{1}}k_{i_{1}/j_{2}}k_{j_{1}/i_{2}}+\sum_{i_{2}j_{1}}k_{i_{2}/j_{1}}k_{j_{1}/i_{2}}+\sum_{i_{1}j_{2}}k_{i_{1}/j_{2}}k_{j_{2}/i_{1}}\label{eq:4}
\end{equation}

Let's note that $\sum_{ij}k_{i/.}k_{j/.}=\sum_{i}k_{i/.}\sum_{j}k_{j/.}$
where the dot may take any value in ${i_{1},i_{2},j_{1},j_{2}}$.

Let $c$ be a community, in equation \prettyref{eq:3-1} summations
$\sum_{ij}k_{i}k_{j}$ on indices $i$ and $j$ may only be applied
under the condition $\delta(c_{i},c_{j})=1$. Where an edge is present
between two nodes $u$ and $v$ belonging to $c$: $\delta(c_{i},c_{j})=1$
and $\delta(c_{j},c_{i})=1$. Consequently for each row $i$ representing
a node belonging to $c$, a corresponding column $j$ represents this
same node belonging to $c$ and \textit{vice versa}. 

From \prettyref{eq:2b-1}, \prettyref{eq:2b-1-1} and the above observations: 

$\sum_{ij}k_{i/j_{1}}k_{j/i_{1}}\delta(c_{i},c_{j})=\sum_{i}k_{i/j_{1}}\sum_{j}k_{j/i_{1}}\delta(c_{i},c_{j})=\sum_{i_{2}}k_{i_{2}/j_{1}}\sum_{j_{2}}k_{j_{2}/i_{1}}\delta(c_{i_{2}},c_{j_{2}})=\sum_{j_{b}}k_{j_{b}}\sum_{j_{b}}k_{j_{b}}=[\sum_{j_{b}}k_{j_{b}}]{{}^2}$

$\sum_{ij}k_{i/j_{2}}k_{j/i_{2}}\delta(c_{i},c_{j})=\sum_{i}k_{i/j_{2}}\sum_{j}k_{j/i_{2}}\delta(c_{i},c_{j})=\sum_{i_{1}}k_{i_{1}/j_{2}}\sum_{j_{1}}k_{j_{1}/i_{2}}\delta(c_{i_{2}},c_{j_{2}})=\sum_{i_{b}}k_{i_{b}}\sum_{i_{b}}k_{i_{b}}=[\sum_{i_{b}}k_{i_{b}}]{{}^2}$

$\sum_{ij}k_{i/j_{1}}k_{j/i_{2}}\delta(c_{i},c_{j})=\sum_{i}k_{i/j_{1}}\sum_{j}k_{j/i_{2}}\delta(c_{i},c_{j})=\sum_{i_{2}}k_{i_{2}/j_{1}}\sum_{j_{1}}k_{j_{1}/i_{2}}\delta(c_{i_{2}},c_{j_{1}})=\sum_{j_{b}}k_{j_{b}}\sum_{i_{b}}k_{i_{b}}$

$\sum_{ij}k_{i/j_{2}}k_{j/i_{1}}\delta(c_{i},c_{j})=\sum_{i}k_{i/j_{2}}\sum_{j}k_{j/i_{1}}\delta(c_{i},c_{j})=\sum_{i_{1}}k_{i_{1}/j_{2}}\sum_{j_{2}}k_{j_{2}/i_{1}}\delta(c_{i_{2}},c_{j_{1}})=\sum_{i_{b}}k_{i_{b}}\sum_{j_{b}}k_{j_{b}}$

where $j_{b}=i_{2}\lyxmathsym{\textendash}r=j_{2}\lyxmathsym{\textendash}r$
, $i_{b}=i_{1}=j_{1}$, $u_{i_{b}}\in c$ and $v_{i_{b}}\in c$ these
last two conditions can also be formalized with $\delta(c_{i_{b}},c_{j_{b}})=1$
if $u_{i_{b}}$ and $v_{i_{b}}$ belong to the same community $c$
and $\delta(c_{i_{b}},c_{j_{b}})=0$ otherwise.

This development yields : 

$\sum_{ij}k_{i}k_{j}=[\sum_{j_{b}}k_{j_{b}}]{{}^2}+[\sum_{i_{b}}k_{i_{b}}]{{}^2}+2[\sum_{j_{b}}k_{j_{b}}][\sum_{i_{b}}k_{i_{b}}]=\sum_{i_{b}j_{b}}(k_{i_{b}}+k_{j_{b}}){{}^2}$
and:
\begin{equation}
\sum_{ij}k_{i}k_{j}\delta(c_{i},c_{j})=\sum_{i_{b}j_{b}}(k_{i_{b}}+k_{j_{b}})^{2}\delta(c_{i_{b}},c_{j_{b}})\label{eq:8}
\end{equation}

Equation \prettyref{eq:8} can be rewritten using the degrees of nodes:

$\sum_{i_{b}}k_{i_{b}}$ is the sum of the degrees of nodes $u_{i_{b}}$
belonging to $c$ under the condition $\delta$ in equation \prettyref{eq:8}.
We denote this $d_{u|c}$ .

$\sum_{j_{b}}k_{j_{b}}$is the sum of the degrees of nodes $v_{j_{b}}$
belonging to $c$ under the condition $\delta$ in equation \prettyref{eq:8}
and has been called $d_{v|c}$.

\begin{equation}
Then\:\sum_{ij}k_{i}k_{j}\delta(c_{i},c_{j})=(d_{u|c}+d_{v|c})^{2}
\end{equation}

\subsection{Analyzing first part in equation \prettyref{eq:3-1}\label{sub:Analysing-first-part}}

First part in $Q$ is $\sum_{ij}A_{ij}^{'}$ . Let's examine what
it represents in terms of $B$. It is possible to identify matrix
$B$ in $A^{'}$ using indices $i_{1}$ and $j_{2}$. Conversely $B^{t}$
can be identified with indices $i_{2}$ and $j_{1}$: 

For $i=i_{1}\: A_{ij}^{'}$s only produce values for $j=j_{2}$, moreover
for $i=i_{2}$, $A_{ij}^{'}$s only produce values for $j=j_{1}$
with $A_{i_{1}j_{2}}^{'}=B_{i_{b}j_{b}}$ and $A_{i_{2}j_{1}}^{'}=B_{i_{b}j_{b}}^{t}$
under typical conditions regarding indices.

Then $\sum_{ij}A_{ij}^{'}=\sum_{i_{1}j_{2}}A_{i_{1}j_{2}}^{'}+\sum_{i_{2}j_{1}}A_{i_{2}j_{1}}^{'}$ 

And $\sum_{ij}A_{ij}^{'}\delta(c_{i},c_{j})=\sum_{i_{1}j_{2}}A_{i_{1}j_{2}}^{'}\delta(c_{i_{1}},c_{j_{2}})+\sum_{i_{2}j_{1}}A_{i_{2}j_{1}}^{'}\delta(c_{i_{2}},c_{j_{1}})$ 

The left-hand side of the sum equals the number of edges from nodes
$u$ to nodes $v$ inside $c$. The right-hand side is the number
of edges from these same nodes $v$ and $u$ inside $c$. This set-up
then leads to: 

$\sum_{i_{1}j_{2}}A_{i_{1}j_{2}}^{'}\delta(c_{i_{1}},c_{j_{2}})=\sum_{i_{2}j_{1}}A_{i_{2}j_{1}}^{'}\delta(c_{i_{2}},c_{j_{1}})\: with\: i_{1}=j_{2}\: and\: i_{2}=j_{1}$

\begin{equation}
Then\:\sum_{ij}A_{ij}^{'}\delta(c_{i},c_{j})=2\sum_{i_{1}j_{2}}A_{i_{1}j_{2}}^{'}\delta(c_{i_{1}},c_{j_{2}})=2\sum_{i_{b}j_{b}}B_{i_{b}j_{b}}\delta(c_{i_{b}},c_{j_{b}})
\end{equation}

This value can also be formalized using the number of edges: 

\begin{equation}
\sum_{i_{b}j_{b}}B_{i_{b}j_{b}}\delta(c_{i_{b}},c_{j_{b}})=|(u_{i_{b|c}},v_{j_{b|c}})|=|e_{i_{b|c},j_{b|c}}|\: where\: e_{i_{b|c},j_{b|c}}\in E\:\&\: u_{i_{b|c}},v_{j_{b|c}}\in c
\end{equation}

For the entire matrix $A^{'}:\sum_{ij}A_{ij}^{'}=2\sum_{i_{b}j_{b}}B_{i_{b}j_{b}}$

From equation \prettyref{eq:3-1}, $m=1/2\sum_{ij}A_{ij}^{'}$ 

Let's now define $m_{b}=\sum_{i_{b}j_{b}}B_{i_{b}j_{b}}=|e_{i_{b}j_{b}}|$
where $e_{i_{b}j_{b}}\in E$ 

Then $m=\frac{1}{2}\times\sum_{ij}A_{ij}^{'}=\frac{1}{2}\times2\times\sum_{i_{b}j_{b}}B_{i_{b}j_{b}}=m_{b}$

\subsection{Modularity for all graphs}

Lastly, by removing sub-index $b$, which had only been introduced
to distinguish indices $i$ and $j$ when applied to $A^{'}$ or $B$,
we can redefine the $A^{'}$ modularity in terms of $B$:

\begin{equation}
Q^{B}=\frac{1}{m}\sum_{ij}[B_{ij}-\frac{(k_{i}+k_{j})\text{\texttwosuperior}}{4m}]\delta(c_{i},c_{j})\label{eq:9-2}
\end{equation}

In terms of edges, by simplifying $e_{i_{b|c},j_{b|c}}$ as $e_{c}$
(where$e_{c}$ has both ends in $c$) and by dropping sub-index $b$
Equation \prettyref{eq:9-2} becomes: 

\begin{equation}
Q^{B}=\sum_{c}[\frac{|e_{c}|}{m}\text{\textendash}(\frac{(d_{u|c}+d_{v|c})}{2\times m})\text{\texttwosuperior}]\label{eq:10}
\end{equation}

This definition of modularity may be used for bipartite graphs since
both types of nodes are bound. In previous sections, we have validated
the above results on the basis of another author's graph modularity
models. It can thus be concluded that equation \prettyref{eq:9-2}
offers a good candidate for bipartite graph modularity that takes
some specific characteristics into account.

\section{Annex 2: Reassignment Modularity function\label{sec:Annex-2}}

In this Appendix, we will provide full details of the demonstration
that yielded Equation \ref{eq:11}.

Reassigning node $w$ from $C_{1}$ to $C_{2}$ either increases or
decreases the modularity defined in Equation \prettyref{eq:9}. Such
a change is referred to as Reassignment Modularity ($RM_{w:C_{1}\rightarrow C_{2}}$).

Let $w$ be a node $u$ or $v$. If $w$ is withdrawn from $C_{1}$
and reassigned to $C_{2}$, then we can define $RM_{w:C_{1}\rightarrow C_{2}}$
=$Q_{w\in C_{2}}^{B}$-$Q_{w\in C_{1}}^{B}$

where $Q^{B}$ is the modularity value in: 
\begin{equation}
Q^{B}=\sum_{c}[\frac{|e_{c}|}{m}\text{\textendash}(\frac{(d_{u|c}+d_{v|c})}{2\times m})\text{\texttwosuperior}].\label{eq:10-1-1}
\end{equation}

Let $l_{w|i}=l_{w,w'|w'\in C_{i}}$ be the number of edges between
a node $w$ and all other nodes $w'$ where $w'\in C_{i}$, 

Let $d_{w}$ be the degree of $w$, \textbar{}$e_{i}|$ the number
of edges in $C_{i}$ and $d_{C_{i}}$= $d_{u|c_{i}}+d_{v|c_{i}}$.

We consider that the node $w$ which belongs to $C_{1}$ is bound
to be withdrawn from this community and assigned to the community
$C_{2}$.

$Q_{w\in C_{2}}^{B}$ is $Q_{w\in C_{1}}^{B}$ with correction after
$w$ is reassigned. Then

$Q_{w\in C_{1}}^{B}$ = $[\frac{1}{m}|e_{1}|-\frac{(d_{C_{1}})\text{\texttwosuperior}}{(2m)^{2}}+\frac{1}{m}|e_{2}|-(\frac{(d_{C_{2}})\text{\texttwosuperior}}{(2m)^{2}})]$
+ $K_{others}$ where $K_{others}$ is the contribution to modularity
brought by other communities than $C_{1}$and $C_{2}$. This last
value does not change when reassigning a node from $C_{1}$ to $C_{2}$.

$Q_{w\in C_{2}}^{B}$ = $[\frac{1}{m}(|e_{1}|-l_{w|1})+\frac{1}{m}(|e_{2}|+l_{w|2})-(\frac{(d_{C_{1}}-d_{w})^{2}}{(2m)^{2}}+\frac{(d_{C_{2}}+d_{w})^{2}}{(2m)^{2}})]$
+ $K_{others}$, then 

$Q_{w\in C_{2}}^{B}$-$Q_{w\in C_{1}}$ = $[\frac{1}{m}(|e_{1}|-l_{w|1})+\frac{1}{m}(|e_{2}|+l_{w|2})-(\frac{(d_{C_{1}}-d_{w})^{2}}{(2m)^{2}}+\frac{(d_{C_{2}}+d_{w})^{2}}{(2m)^{2}})]-[\frac{1}{m}|e_{1}|-\frac{(d_{C_{1}})\text{\texttwosuperior}}{(2m)^{2}}+\frac{1}{m}|e_{2}|-(\frac{(d_{C_{2}})\text{\texttwosuperior}}{(2m)^{2}})]$

and after simplification, 
\begin{equation}
RM_{w:C_{1}\rightarrow C_{2}}=\frac{1}{m}(l_{w|2}-l_{w|1})-\frac{1}{2m{}^{2}}[d_{w}^{2}+d_{w}(d_{C_{2}}-d_{C_{1}})]\label{eq:11-1}
\end{equation}

This equation can be partly validated if after withdrawing $w$ from
$C_{1}$ we put it back into $C_{1}$ and expect no change for $Q^{B}$,
i.e. $RM_{w:C_{1}\rightarrow C_{1}}=0$. Considering that $C_{2}$
is in fact $C_{1}$ without $w$, we get $d_{C_{2}}=d_{C_{1}}-d_{w}$,
replacing $d_{C_{2}}$ in equation \prettyref{eq:11-1} by its value
yields $RM_{w:C_{1}\rightarrow C_{1}}=0$.

A second validation can be performed with Equation 5 in \cite{Wu2012a}.
Although the authors' demonstration is limited, it can still be noticed
that their final formula resembles ours with a slight difference (i.e.
division by 2 in their case) due to their definition of modularity
for overlapping communities. Moreover, in arguing that the right part
of their equation is not meaningful for large graphs, the authors
only considered $dEQ=\frac{l_{2}-l_{1}}{2m}$ which is the equivalent
of $\frac{1}{m}(l_{w|2}-l_{w|1})$ in our Reassignment Modularity
definition. In our case, we do not limit reassignment to large graphs
and we keep the whole value in Equation \prettyref{eq:11-1}. 
\end{document}